\documentclass[preprint]{elsarticle}
\usepackage{epstopdf}
\usepackage{amssymb}
\journal{Astroparticle Physics}

\begin{document}

\begin{frontmatter}

\title{In-situ measurements of whole-dish reflectivity for VERITAS}


\author{S.~Archambault}
\author{G.~Chernitsky}
\author{S.~Griffin}
\author{and D.~Hanna\corref{cor1}}

\cortext[cor1]{Corresponding author: hanna@physics.mcgill.ca}

\address{Physics Department, McGill University, Montreal, QC H3A 2T8, Canada}

\begin{abstract}
{The VERITAS array is a set of four imaging atmospheric Cherenkov telescopes 
(IACTs) sensitive to gamma rays with energies above 80 GeV.
Each telescope is based on a tessellated mirror, 12 metres in diameter, 
which reflects light from a gamma-ray-induced air shower to form an image 
on a pixellated `camera' comprising 499 photomultiplier tubes. 
The image brightness is the primary measure of the gamma ray's energy so a 
knowledge of the mirror reflectivity is important. 
We describe here a method, 
pioneered by members of the MAGIC collaboration, to measure the whole-dish 
reflectivity, quickly and regularly, so that effects of mirror 
aging can be monitored.
A CCD camera attached near the centre of the dish simultaneously acquires an 
image of both a star and its reflection on a target made 
of Spectralon, a 
highly-reflective material, placed at the focus of the telescope. The ratio of 
their brightnesses, as recorded by the CCD, along with geometric factors, 
provides an estimate of the dish reflectivity with few systematic errors. 
A filter wheel is deployed with the CCD camera, allowing one to measure the 
reflectivity as a function of wavelength. We present results obtained 
with the VERITAS telescopes since 2014.}

\end{abstract}


\end{frontmatter}
\section{Introduction}
 
Very-high-energy (VHE) gamma-ray astronomy makes use of arrays of imaging
atmospheric Cherenkov telescopes (IACTs).
A telescope consists of a large tesseleted mirror that focusses Cherenkov light 
from extensive air showers onto a camera comprising an array of photomultiplier
tubes (PMTs).
The sum of signals from the PMTs is proportional to the energy of the  
incident gamma ray, so a key parameter needed to 
extract science from the data acquired using such 
telescopes is the effective reflectivity of the mirror, the
``whole-dish'' reflectivity.
This number can be estimated using measurements on 
a representative set of 
facets made periodically with a laboratory setup.
Calculated 
corrections to account for the shadowing effects of the PMT camera and its
support structure can also be included.
However it is desirable to have alternative ways of determining the 
mirror reflectivity if only to build confidence in one's understanding of the
instrument and as a way of estimating systematic errors.

In this report we describe our experience with a system developed to measure
the whole-dish reflectivity of the VERITAS telescopes. 
The method we follow was first suggested by members of the MAGIC 
collaboration~\cite{magic1,magic2, magic3}.
The basic idea is to mount a digital camera on the telescope to
record, in the same image, 
light coming directly from a bright star as well as the light
from that star that has reflected off the main mirror and subsequently off a
target of known reflectivity placed at the mirror focus. 
Up to numerical factors, the whole-dish reflectivity is determined from 
the ratio of the two signals.
The use of a single camera to simultaneously record both the direct and 
reflected images eliminates many possible systematic errors.

\section{Apparatus}

VERITAS comprises an array of four IACTs located at the Whipple Observatory 
on Mount Hopkins in southern Arizona~\cite{park, holder}.
Each of the telescopes is based on a 12-m diameter Davies-Cotton reflector 
focussing light onto a 499-pixel camera made from close-packed 
Hamamatsu R10560 PMTs coupled to light concentrators.  
The reflector is made up of 345 identical mirror facets; when they are 
perfectly aligned 
the on-axis point-spread-function is smaller than a pixel 
diameter~\cite{mccann}.

\subsection{Digital Camera}

The digital camera used for this work is the ST402ME model from Santa Barbara
Instrument Group (SBIG - now Diffraction Limited\textcircled{c}), 
an  ``entry-level'' model which nevertheless 
incorporates some fairly advanced features.
Given that we wanted to equip all four telescopes in the VERITAS array with 
its own permanently-mounted camera to allow parallel data taking and 
stability between measurement sessions, the relatively low 
price was a factor in its choice.

The camera is based on a Kodak KAF-0402ME CCD chip with 765 $\times$ 
512 pixels and  85\% peak quantum efficiency. 
An important feature of this camera is its on-board thermo-electric 
cooling system which can 
cool the CCD chip to approximately 25 C below ambient temperature, thus 
lowering the dark current considerably.
Automatic dark-frame subtraction is available to obviate the effects of 
stuck or hot pixels.

The lens is a Ricoh C32500 machine vision lens with 25 mm focal length and 2/3''
format. It is used with maximum iris opening (f/1.4) and is focussed on the
Spectralon at a distance of 12 m. 

\begin{figure}[]
\centering
\includegraphics[width=0.90\textwidth]{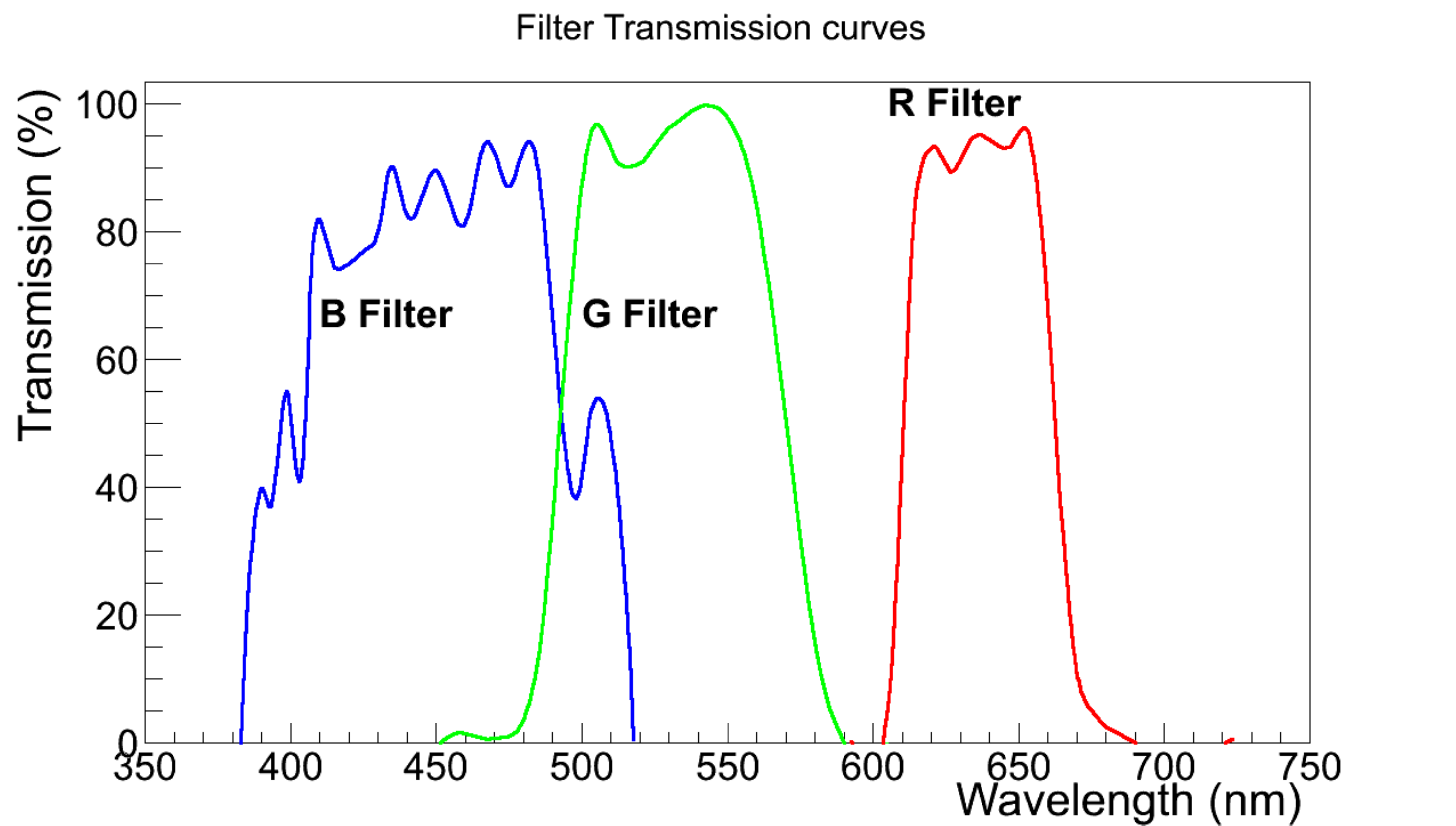}
\caption{Transmission curves of the RGB filter system. 
(Data supplied by SBIG/Diffraction Limited\textcircled{c}).}
\label{filter}
\end{figure}

The camera is equipped with an integrated RGB filter wheel to allow measurements
at different wavelengths.
The transmission bands, shown in Figure~\ref{filter}, 
are rather wide so wavelength-dependence 
investigations are consequently approximate. 
The use of third-party filters with narrower transmission windows is precluded 
by the small size of the filter wheel; the filters provided by SBIG are 
custom-made to fit.

The camera is controlled and read out by a 
small form-factor computer built from 
the following commercially available components:

\begin{itemize}

\item{Motherboard with integrated CPU: ASRock E350M1 AMD E-350 APU}

\item{Power supply: APEVIA ITX-AP250W}

\item{Hard drive: ADATA SP900 ASP900S3-64GB-C 64GB SSD}

\end{itemize}

The computer uses the Ubuntu 12.04 LTS operating system and has an ethernet 
connection for control signals and data retrieval.

\begin{figure}[]
\centering
\includegraphics[width=0.90\textwidth]{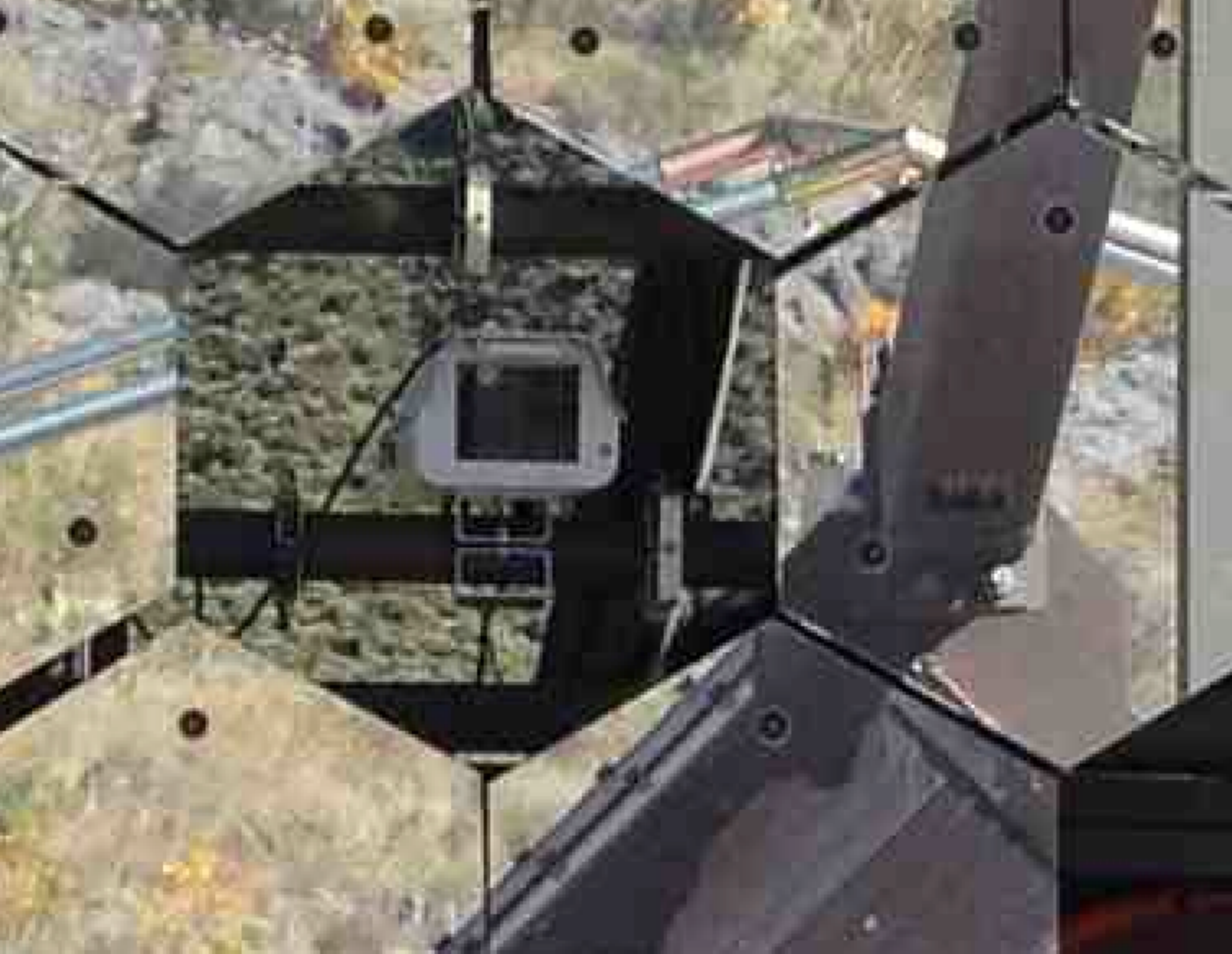}
\caption{A photograph of the Pelco box containing the SBIG camera and 
control computer mounted near the centre of a VERITAS reflector, 
in place of one of the mirror facets.
}
\label{pelco}
\end{figure}

The computer and camera are contained in a weatherproof box (Pelco model 
EH5700) mounted on 
the optical support structure of the telescope, approximately 
1.7 m from the centre of the dish, as shown in Figure~\ref{pelco}.

\subsection{Reflective Target}

\begin{figure}[]
\centering
\includegraphics[width=0.90\textwidth]{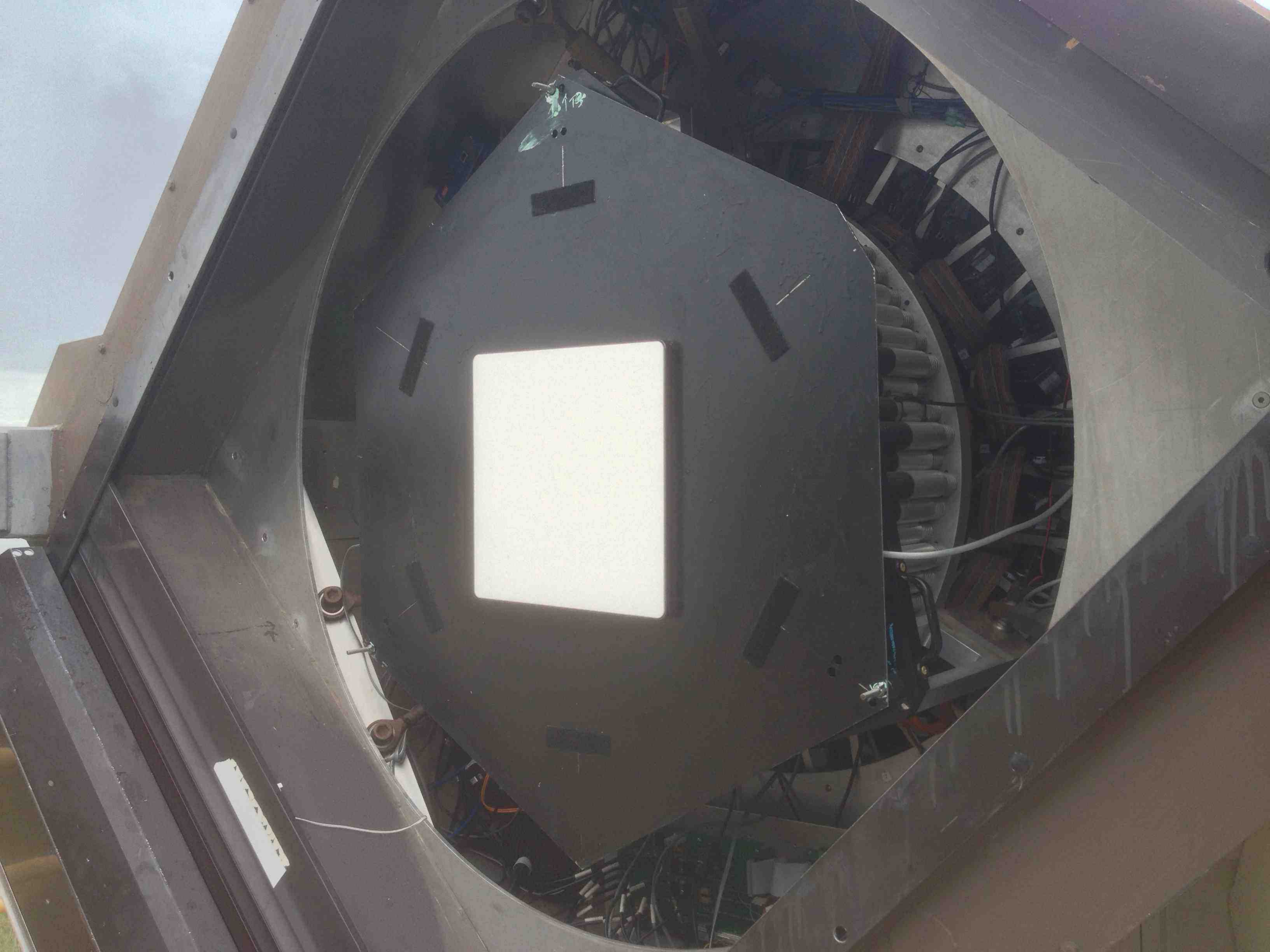}
\caption{The Spectralon target and its mounting plate is shown 
as installed on one of the VERITAS cameras, just in front of the PMTs.}
\label{target}
\end{figure}

The secondary reflector onto which starlight is focussed is a square piece of
Spectralon, supplied by Labsphere Inc., 30 cm on a side.
It is a fluoropolymer with diffuse (Lambertian) 
reflectance greater than 99\% over the wavelength range of interest.
The target is attached to an aluminum plate that can be temporarily mounted
at the focal point of the telescope, as shown in 
Figure~\ref{target}. 
When not in use, the target is stored with a protective cover to prevent 
any dirt or dust from reducing its reflectivity.
The targets were returned to the manufacturer for recalibration two years after
purchase and showed no significant deterioration.
In 2020, a new target was purchased and used for a set of measurements
on one of the VERITAS telescopes, as a check.
Reflectivity values for the telescope 
were consistent with those obtained with the old target.
We conclude that any apparent drop in the dish reflectivity is due to the 
mirror facets and not to any changes in the Spectralon targets.

\section{Data Acquisition}

\begin{figure}[]
\centering
\includegraphics[width=0.90\textwidth]{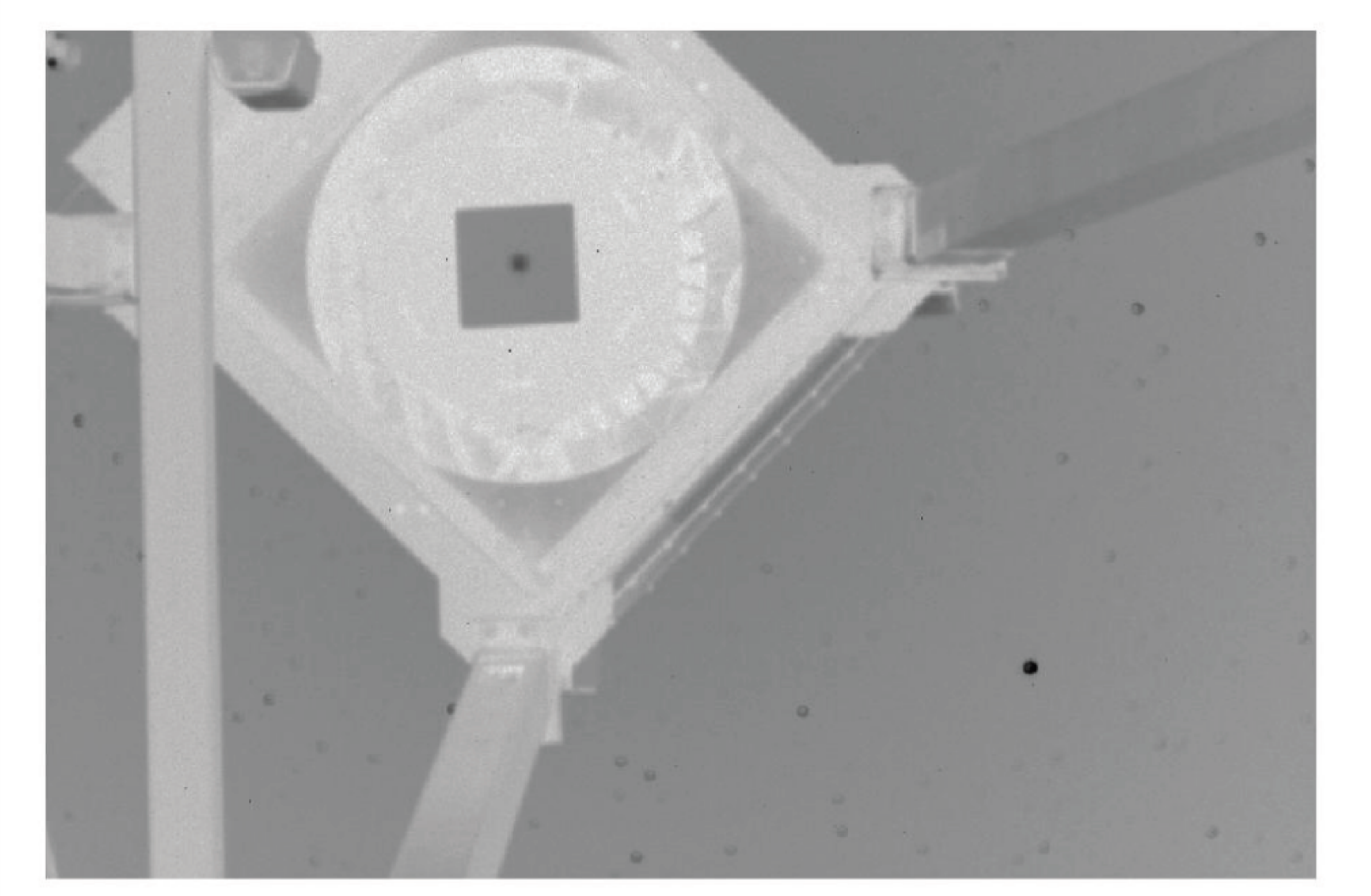}
\caption{A sample image of the kind used in the reflectivity measurements.
The scale is logarithmic and the grey-scale is inverted to highlight the 
target star (the black object in the lower right quadrant)
and its reflection on the Spectralon (upper left).
}
\label{fits}
\end{figure}

Figure~\ref{fits} shows a sample of the type of image used in the reflectivity
measurements.
A logarithmic intensity scale and inverted grey-scale highlight the key 
features.
One sees an intense black object in the lower right quadrant; this is the 
direct image of the target star.
Its finite extent is due to its being out of focus because the CCD camera 
is focussed on the Spectralon target seen in the upper left quadrant.
This out-of-focus feature allows one to avoid saturation when dealing with 
bright stars or long exposures, both needed for improving signal to noise.

The telescope focus box and supporting quadrupod arms are clearly visible, 
as is the square shape of the Spectralon target.
The reflection of the starlight focussed thereon is seen as a round black 
object in the middle.
Its size is largely due to the PSF of the reflector dish.

The reflectivity measurement protocol involves the acquisition of images from 
four or five bright stars at elevations greater than
50$^\circ$.
The angle cut reduces any atmospheric effects and anthropogenic
backgrounds.
The images are usually obtained when there is partial moonlight in order not to
interfere with gamma-ray observing. 
For each star, an unfiltered image is acquired and the exposure is checked 
for saturation, with adjustments to the exposure time being made, 
automatically, if necessary.
Quasi real time feedback is supplied to the operator via SAOImage 
DS9~\cite{DS9}. 
The first image is obtained without dark-frame subtraction activated to allow
an easy check for saturation.
Next, two or more additional images are obtained with the subtraction 
feature switched on. 
This procedure is repeated with the R, G, and B filters in the optical 
path.
All data are stored as FITS files on the local disk and are later transferred  
to an archive.

All four telescopes are equipped with identical apparatus 
and the entire procedure
takes on the order of thirty minutes for a complete data set.

\section{Data Analysis}

\subsection{Signal extraction}
The off-line analysis proceeds as follows. 
The dark-frame-subtracted files are combined and the centres of the stars
and their reflections are determined.
Concentric squares are positioned around the centre, an inner one typically 
24 $\times$ 24 pixels and an outer one 36 $\times$ 36 pixels.
The sum of the pixel values from the inner square, minus the 
geometrically-scaled sum of pixel values from the region 
between the inner and outer 
square boundaries, used as 
an estimate of the background, constitutes the signal.
The results are independent of the precise values of the square sizes,
as long as the reflection image is fully contained within the inner square.

The whole-dish reflectivity $R$ is calculated from the direct and reflected signals, $S_d$ 
and $S_r$, respectively, using the 
formula $ R = (S_r / S_d) \pi d^2 / A_M $ where 
$A_M$ is the area of the dish (115 m$^2$) and $d$ is the distance from 
the Spectralon to the CCD camera (12 m).
This assumes that the reflectivity of the Spectralon is 100\% and that the 
angle 
between the CCD camera axis and the normal to the Spectralon target is 
approximately zero, so that there is no cosine-dependent correction for
the Lambertian distribution of light coming from the target. 
Note that this formula for $R$ is for the {\it effective} reflectivity of the 
dish.
It includes the effects of any missing or misaligned mirror facets or of 
shadowing of the PMT-based camera due to the camera itself and its support 
structure. 
Both of these result in a reduction of light reflecting off the dish and onto 
the Spectralon, with a consequent reduction in $S_r$ but not $S_d$. 
$R$ is the parameter of most use in the analysis of gamma-ray data.
It is agnostic as to the reason light was lost, whether it be to degradation
of the facet reflectivity, misalignment, or shadowing.

\subsection{Vignetting correction}

\begin{figure}[]
\vspace{-1.5cm}
\centering
\includegraphics[width=1.0\textwidth]{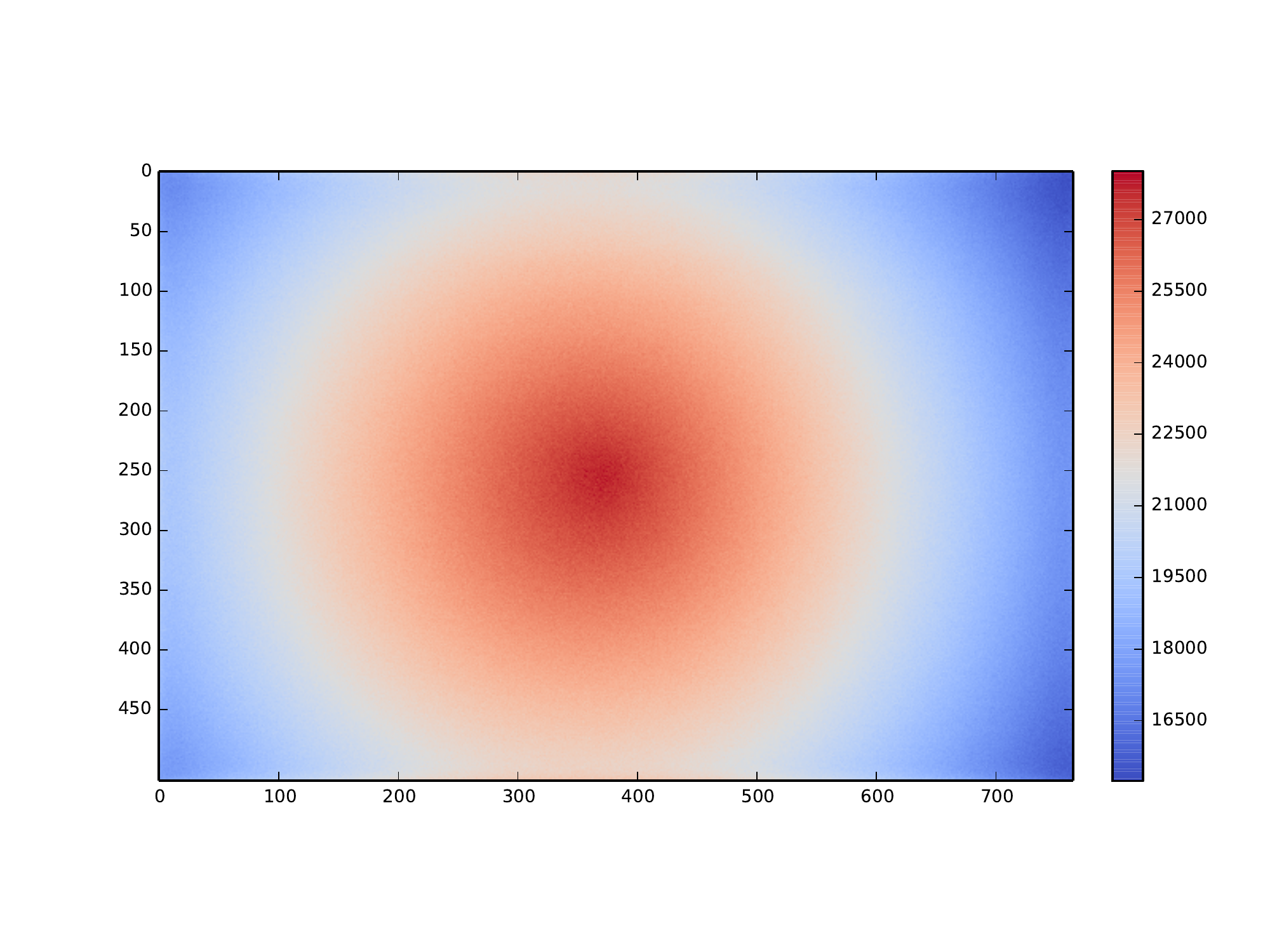}
\vspace{-1.5cm}
\caption{A sample flat-field image taken with a camera pointing to the 
zenith at twilight.
The range of pixel values is almost a factor of two.
Information from images like this  
is used to correct for vignetting effects in the 
reflectivity images.
}
\label{flatfield}
\end{figure}

The CCD camera produces images that suffer from vignetting effects and this 
must be accounted for in calculating the reflectivity of the mirror.
To do this we use flat-field images obtained by photographing the zenith at 
twilight using the same focus and aperture settings as with normal data.
A typical image is shown in Figure~\ref{flatfield}. 
The difference in intensities between the centre of the image and the corners 
is almost a factor of two, indicating the importance of corrections.
The data are well parameterized by a conical fit.
The fit parameters are the centre of cone and a single slope which describes the
fall-off of intensity with radius from the centre.
The inverse of this function is used to unfold the vignetting effects.  

The correction factors can be reduced by careful aiming of the camera.
With the star image and the reflection image approximately equidistant from 
the centre of the camera's field-of-view, each suffers approximately the same
amount of vignetting. 
Since we use the ratio of measured intensities, the 
correction amounts to a second-order effect.

\subsection{Statistical uncertainties}

The data from Figure~\ref{flatfield} are useful for determining statistical 
uncertainties.
We assume that adjacent pixels should report approximately 
the same number in a flat-field
image since the twilight sky is uniform and vignetting effects vary slowly 
across the image plane.
The largest contribution to any difference between adjacent values should be due
to statistical fluctuations. 
Thus we can histogram the differences between values from neighbouring pixels
in a row (or column) of the CCD
and use the width of the resulting distribution as an estimator
of the statistical error.
The effect varies, as expected, 
with the magnitudes of the pixel values and can be parameterized as
$\sigma^2_q \simeq 0.85 q$, where $q$ is the pixel value.
This dependence is used in assigning uncertainties to pixel values in the 
analysis.

\section{Results}

In this section we present results obtained since January, 
2014, when development of the measurement system was completed.

\begin{figure}[]
\centering
\includegraphics[width=0.90\textwidth]{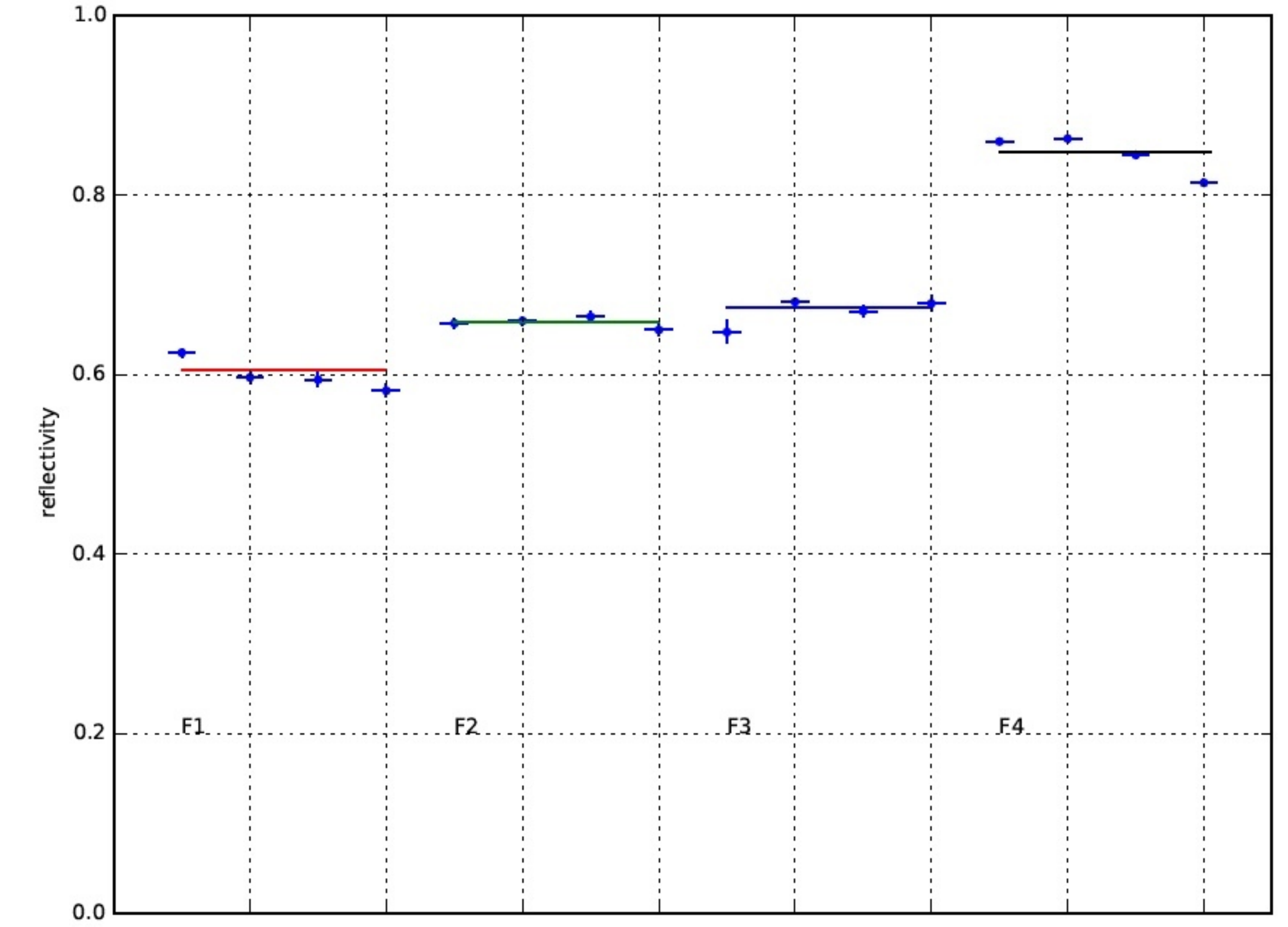}
\caption{
Results from a single telescope for a single observing session.
The reflectivities, measured using four different stars and four 
band-pass filters, are plotted. 
The points are grouped according to filter colour (red, green, blue,
and clear) and a flat line, 
representing the average value, is superimposed.
Uncertainties are statistical.
}
\label{t2-general}
\end{figure}

Figure~\ref{t2-general} shows a sample of results from a single observing 
night made with one of the VERITAS telescopes.
Reflectivities, measured using four different stars and four band-pass 
filters, are displayed, with results grouped by filter colour.
A flat line indicates the weighted mean for each group. 
Statistical uncertainties are plotted.

Similar plots exist for the three other telescopes in the array and for 
different observing sessions.
In all cases the reflectivity values are highest for the blue-filter data 
and lowest for the red-filter data, consistent with the canonical 
wavelength dependence for aluminum.
Values obtained without filters show more scatter due to the effects of the
different colours of the stars used.

The average values for each filter and each data set can be plotted vs time
to show the evolution of the reflectivity with time.
Here we show results from blue-filter data, the most relevant for Cherenkov
telescopes since most of the detected light is from the UV and blue parts of the
spectrum. 
(Note that the CCD camera is insensitive to UV wavelengths;
there is a sharp cutoff in quantum efficiency at 400 nm.)

\begin{figure}[]
\centering
\includegraphics[width=1.0\textwidth]{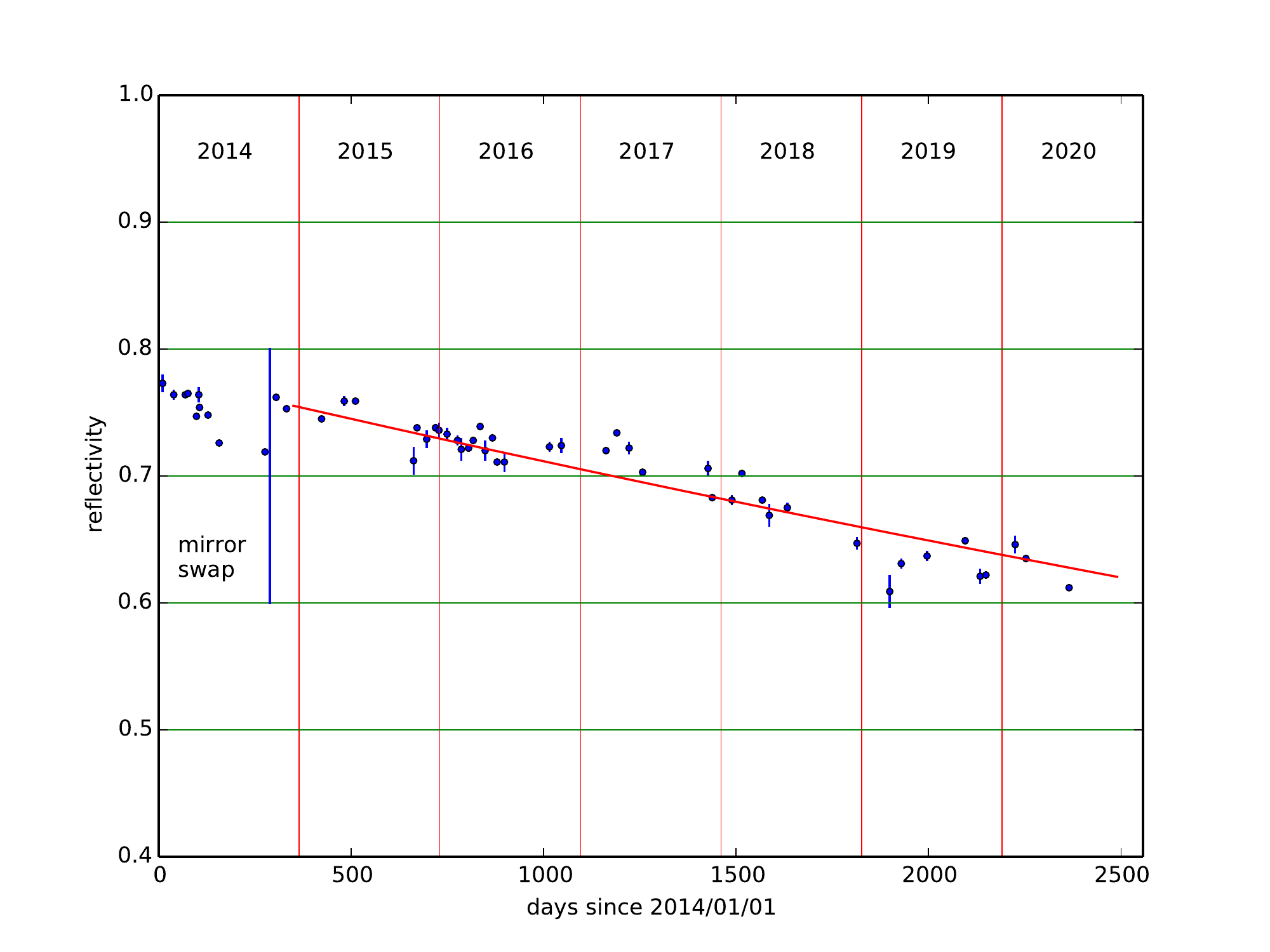}
\caption{Time evolution of the reflectivity at blue wavelengths for one of the 
VERITAS telescopes. A step up 
in reflectivity near day 260 is due to a mirror swap 
where a third of the facets were replaced with freshly surfaced ones.
Other scatter in the data is due to residual systematic errors.
The curve is an exponential drawn to emphasize the nature of the decline and 
indicate the level of systematic uncertainties.}
\label{t2_blue}
\end{figure}

In Figure ~\ref{t2_blue}, 
we have results from one of the VERITAS telescopes plotted vs days since the 
begining of 2014.
There is a clear secular decline caused by weathering of the mirror coatings.
Monthly washing of the mirrors helps to keep the effects of dust and dirt from
making the decline more rapid and may contribute to some of the scatter seen
in these plots.
The decline is gradual but significant; the red line 
is not a fit but is drawn to show that the 
decline is approximately exponential and to indicate the level of systematic
uncertainties that remain.
All four VERITAS telescopes show similar long-term behaviour. 

The scatter of points about the line indicates the level of systematic 
uncertainties. 
Chief among them is the effect of vignetting.
This can be seen in the differences from season to season that are not 
fully corrected. 
This results from taking down the cameras during summer shut-downs; 
with the current setup it is difficult to aim them precisely when they are 
reinstalled at the start of the observing season.
A better mounting system would help with this situation.

VERITAS has approximately 100 spare mirror facets and these are used in a 
program wherein facets are taken off the telescopes 
and replaced by the spare facets~\cite{roache}. 
The replaced facets are then recoated and used as replacements in the 
next facet swap.
A facet swap was performed on this telescope just after day 250 
and the effect of
this operation can be seen in the figure.

\section{Conclusion}
We have successfully implemented a system to measure the reflectivity
of the VERITAS telescope mirrors.
The measurements can be carried out in less than one hour and at times 
that do not interfere with 
gamma-ray observations (late twilight or under partial moonlight).
The use of band-pass filters to restrict the wavelengths observed lessen any 
confounding effects arising from the use of stars of different spectral
type and ensure that reflectivities obtained from different stars are 
consistent within errors.
Our observations over seven years indicate that the system is 
relatively stable and is able to track mirror degradation at a level of a few
percent.
Systematic errors are mainly due to vignetting effects and could be reduced
by using a camera mounting system that would allow precision adjustment to 
where the target star and its reflection appear in the camera image.

\section{Acknowledgements}
This research is supported by grants from the U.S. Department of Energy Office 
of Science, the U.S. National Science Foundation and the Smithsonian 
Institution, by NSERC in Canada, and by the Helmholtz Association in Germany. 
We acknowledge the excellent work of the technical support staff at the Fred 
Lawrence Whipple Observatory and at the collaborating institutions in the 
construction and operation of the instrument.
We are grateful to our VERITAS colleagues for help with acquiring
the data for the studies described here.
Conversations with R. Mirzoyan have been very helpful.


\end{document}